# Materials genes of heterogeneous catalysis from clean experiments and artificial intelligence

Lucas Foppa,[a,b]* Luca M. Ghiringhelli,[a,b] Frank Girgsdies,[a] Maike Hashagen,[a] Pierre Kube,[a] Michael Hävecker,[c] Spencer J. Carey,[a] Andrey Tarasov,[a] Peter Kraus,[a,†] Frank Rosowski,[d] Robert Schlögl,[a,c] Annette Trunschke,[a] and Matthias Scheffler [a,b]

[a]Fritz-Haber-Institut der Max-Planck-Gesellschaft, Faradayweg 4-6, D-14195 Berlin, Germany. [b]Humboldt-Universität zu Berlin, Zum Großen Windkanal 6, D-12489 Berlin, Germany, [c]Max-Planck-Institut für Chemische Energiekonversion, Stiftstr. 34-36, D-45470 Mülheim, Germany, [d]BASF SE, Process Reseach and Chemical Engineering, Heterogeneous Catalysis, Carl-Bosch-Straße 38, D-67065 Ludwigshafen, Germany.

**Abstract:** Heterogeneous catalysis is an example of a complex materials function, governed by an intricate interplay of several processes, e.g., the different surface chemical reactions, and the dynamic re-structuring of the catalyst material at reaction conditions. Modelling the full catalytic progression via first-principles statistical mechanics is impractical, if not impossible. Instead, we show here how a tailored artificial-intelligence approach can be applied, even to a small number of materials, to model catalysis and determine the key descriptive parameters ("materials genes") reflecting the processes that trigger, facilitate, or hinder catalyst performance. We start from a consistent experimental set of "clean data", containing nine vanadium-based oxidation catalysts. These materials were synthesized, fully characterized, and tested according to standardized protocols. By applying the symbolic-regression SISSO approach, we identify correlations between the few most relevant materials properties and their reactivity. This approach highlights the underlying physicochemical processes, and accelerates catalyst design.

**Introduction**

The identification of physicochemically meaningful, descriptive parameters that are correlated with catalyst performance is a key step for modelling and understanding heterogeneous catalysis and finding new and more efficient catalytically active materials. These parameters, which characterize the materials and the processes triggering, facilitating or hindering the reaction, might be called the *materials genes* of heterogeneous catalysis. These catalyst genes can be used to construct *maps of catalysts, i.e.,* materials charts that highlight the small interesting regions of the (huge) space of all possible materials, where the search for high-performance catalysts should be focused.[1, 2] However, finding such descriptive parameters is challenging because the outcomes of interest (e.g., product selectivity) in reactions catalyzed by solids result from the concerted and intricate interplay of several processes. These are related to the material itself but also to the reaction conditions, for instance the temperature and gas-phase in contact with the solid. Some of these processes are: multiple bond-breaking and -forming reactions occurring on the catalyst surface, the coverage of adsorbates on those surfaces, the catalyst re-structuring in the reaction environment, referred to as the *catalyst dynamics*,[3] and the diffusion of reactants and products within the porous structure of the catalyst.[3, 4]

One approach for describing heterogeneous catalysis is the theoretical, multi-scale modelling by first-principles simulations.[4-6] Nevertheless, the atomistic modelling of the full catalytic progression under realistic conditions is impractical because it requires computationally prohibitive methods for the accurate evaluation of large, interconnected networks of surface reactions[7, 8] and complex statistical-mechanical treatments of the catalyst dynamics.[9] Additionally, mesoscale (e.g., adsorbate-adsorbate) and transport phenomena need to be taken into account as well. Finally, the coupling of all these phenomena, occurring at very different time and length scales is highly complex (see references 4, 5 and references therein). While experiments, for example spectroscopic studies under reaction conditions, can point to the specific processes governing the reactivity on the particular systems under investigation, it is not obvious how to derive general and quantitative relationships between materials physicochemical properties (and reaction conditions) and the catalyst performance that go beyond the classical Sabatier principle of optimal binding strength between reacting species and the catalyst.[4, 10, 11]

In this paper, we demonstrate how a tailored artificial-intelligence (AI) approach, even when applied to only a small number of materials and materials functions, but billions of quantitative materials features, can determine the key physicochemical descriptive parameters characterizing the catalyst performance. This method is used to address the challenging propane selective oxidation reaction. We start from a consistent experimental set of "clean data", containing nine vanadium-based oxidation catalysts (Fig. 1A). Here, the term "clean data" refers to the fact that these materials were carefully synthesized and tested in catalysis according to standardized protocols.[12] Importantly, these nine catalysts were also characterized in detail, resulting in more than forty measured properties per material. To this data set, we applied the compressed-sensing symbolic-regression sure-independence-screening-and-sparsifying-operator (SISSO)[13, 14] approach (Fig. 1B). We thus identified the few most relevant parameters that are correlated, in a possibly complicated way, with the selectivity towards acrylic acid and with catalyst activity.

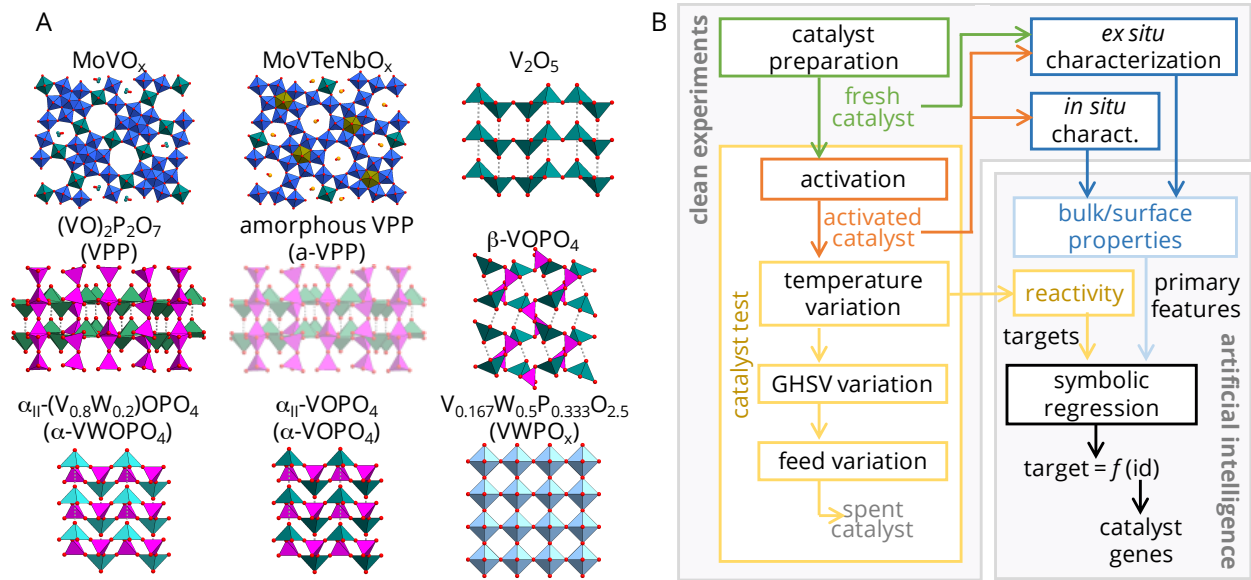

**Figure 1.** (A) Vanadium-based selective oxidation catalysts used in this work. (B) Schematic workflow of the proposed approach combining clean experiments and AI for the identification of "materials genes" of heterogeneous catalysis. Here GHSV means "gas hourly space velocity", and *f(*id) means "function of interpretable descriptors". The "catalyst preparation" step consists in catalyst synthesis, calcining, pressing and sieving.

**Alkane selective oxidation**

The selective oxidation reaction performed with mixed-metal-oxide heterogeneous catalysts enables the transformation of abundant light-alkanes (e.g., ethane, propane and *n*-butane) into the valuable products olefins and oxygenates.[15] However, the initial alkane might undergo multiple reactions on the surface of the catalyst in the presence of oxygen molecules ($O_2$),[7, 16] leading not only to the desired molecules but also to several by-products, including $CO_2$. The chemical equations describing the formation of propylene ($C_3H_6$, olefin), acrylic acid ($C_3H_4O_2$, oxygenate), and $CO_2$ (combustion or total-oxidation product) in propane ($C_3$) oxidation, for instance, are:

$$2\ C_3H_{8(g)} + O_{2(g)} \rightarrow 2\ C_3H_{6(g)} + 2\ H_2O_{(l)}\ (-162\ kJ \cdot mol^{-1})\ \text{(eq. 1)},$$
$$C_3H_{8(g)} + 2\ O_{2(g)} \rightarrow C_3H_4O_{2(l)} + 2\ H_2O_{(l)}(-852\ kJ \cdot mol^{-1})\ \text{(eq. 2)},$$
$$\text{and}$$
$$C_3H_{8(g)} + 5\ O_{2(g)} \rightarrow 3\ CO_2 + 4\ H_2O_{(l)}\ (-2220\ kJ \cdot mol^{-1})\ \text{(eq. 3)}.$$

Here, the values in parenthesis are the standard reaction enthalpies.[17] Selectively forming the desired products, and, in particular the value-added oxygenate, is therefore a challenge. The "seven pillars" of oxidation catalysis indicate the several factors contributing to reactivity in oxidation reactions: 1. lattice oxygen, 2. metal–oxygen bond strength, 3. host structure, 4. redox properties, 5. multifunctionality of active sites, 6. site isolation, and 7. phase cooperation.[18, 19] In the case of vanadium-based oxide catalysts, selectivity has been also related to surface enrichment of one of the metal ions in the presence of reaction feed containing steam[20-22] and the associated surface potential barrier,[20, 23, 24] highlighting that the catalyst dynamics also plays a role. Due to the multiple requirements and the intricacy of the underlying processes, the theoretical description of selective oxidation and the search for new catalysts is extremely challenging. Alternative approaches for modeling and designing new catalysts are thus required. Here, we propose a combination of standardized experiments and AI to address this problem.

**Experimental handbooks for the generation of "clean data"**

The identification of reactivity descriptors by AI relies on the consistency of the input data. Therefore, we developed standardized protocols for catalyst synthesis, characterization and testing, described in *experimental handbooks*,[12] which enable the generation of consistent and annotated data, according to the FAIR principles (Findable, Accessible, Interoperable and Re-purposable/Re-usable).[1] The establishment of minimum requirements for performing and reporting measured reactivity is a crucial aspect in heterogeneous catalysis research. Because kinetic effects play a dominant role in catalysis, the reactivity is not only sensitive to the catalyst synthesis procedure and to the resulting as-synthesized, *fresh* catalyst, but also to the conditions to which the material is exposed prior to and during the reaction, for instance the temperature and the composition of the gas-phase (feed) in contact with the solid.[2]

In this work, we focus on nine common vanadium-based oxidation catalysts (Fig. 1A). These materials were prepared, in a reproducible manner, in large batches (15-20 g) to guarantee that comprehensive catalyst characterization and testing is performed using samples from the same batch. The catalyst preparation consists on the catalyst synthesis itself, plus calcining, pressing and sieving. The materials resulting from the catalyst preparation are called *fresh catalysts*. After catalyst preparation, all the catalysts were tested for the $C_3$-oxidation reaction using a fixed-bed reactor (Fig. 1B). The *catalyst test* starts with an activation procedure during which the synthesized materials are exposed to the reaction feed and rather high temperature (e.g., 450 ºC) for 48 hours. The activation condition is that the conversion of either propane or oxygen is 85%. The materials resulting from the activation procedure are called *activated catalysts*. The goal of the activation procedure is to obtain samples as similar as possible to the catalytically active materials formed during the induction period of the reaction.

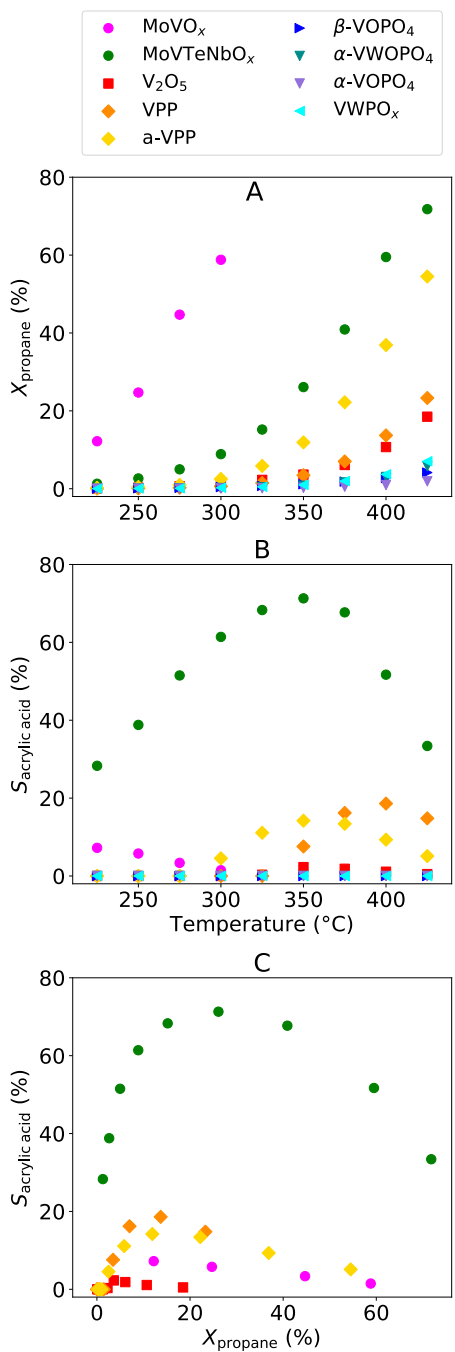

**Figure 2.** (A) Propane conversions ($X_{propane}$) and (B) acrylic acid selectivity ($S_{acrylic\ acid}$) of the vanadium-based catalysts measured in the catalyst test, evidencing the diverse types of behavior in the data set. (C) $S_{acrylic\ acid}$ dependence on $X_{propane}$, showing that acrylic acid, as a consecutive product, is only formed at intermediate conversion levels. The MoVO$_x$ catalyst performance was not measured at temperatures above 300 °C due to its limited thermal stability in the feed. Only the catalysts that produce acrylic acid are shown in (C).

Indeed, some catalysts undergo structural modifications during the activation procedure. For this reason, their properties differ significantly between fresh and activated catalyst samples (see data set provided in Electronic Supporting Information, ESI).

Following the activation step of the catalyst test, the temperature is brought to 225 °C in lean air and gradually increased, in steps of 25 °C, in the reaction feed up to 450 °C, to enable the conversion of propane and oxygen. If the propane and/or the oxygen (molar) conversion exceeds 85%, the increase in temperature is stopped to prevent catalyst decomposition. At each of the temperatures, the steady-state operation is reached and the reaction mixture at the reactor outlet is collected and analyzed, providing the measures of catalytic performance. Catalyst *activity* and *selectivity* are evaluated in terms of propane conversion ($X_{propane}$) and product selectivity ($S_{product}$), respectively. The propane conversion indicates the molar fraction of oxidized propane, i.e. propane converted to any of the possible products. The selectivity indicates the molar fraction of a specific product with respect to all products formed from propane. The gas hourly space velocity (GHSV), the ratio between the volumetric flow and the catalyst volume, is kept constant for all catalysts during the test (at 1000 h$^{-1}$) to ensure a consistent comparison among materials. After the temperature variation, the GHSV and feed are varied and the spent catalyst is further analyzed. These steps beyond temperature variation are not discussed in this paper. The raw data used here are provided as ESI and the detailed experimental procedure is explained in the *handbook*.[12]

The performance of the nine catalysts in C$_3$-oxidation, in terms of propane conversion and acrylic acid selectivity (Fig. 2), shows a wide range of behaviors in the nine selected vanadium-based catalysts. These catalysts have different activity, i.e., they react with propane in different amounts, as indicated by the different propane conversions profiles (Fig. 2A). MoVO$_x$ is much more active than the other catalysts and converts 58.8 % of the initial propane at 300 °C. At higher temperatures, MoVTeNbO$_x$, a-VPP, VPP and V$_2$O$_5$, achieve conversions higher than ca. 20 %, with MoVTeNbO$_x$ being the most active catalyst among them. β-VOPO$_4$, α-VWOPO$_4$, α-VOPO$_4$ and VWPO$_x$ achieve significantly lower conversions (below ca. 10 %), even at the highest applied temperature. These catalysts are therefore the least active materials. Several products are formed from the initial propane on each catalyst, including the value-added acrylic acid and propylene, as well as the undesirable CO$_2$. In our analysis, we focus on acrylic acid because the formation of this product involves a complex interplay of processes. The acrylic acid selectivity measured at each temperature of the catalyst test is shown in Fig. 2B. Acrylic acid is a consecutive product of the propane oxidation reaction. It is formed after propane transformation to propylene, but before the total oxidation product CO$_2$ (see eq. 1-3). For this reason, its formation is only observed at intermediate propane-conversion levels (Fig. 2C). MoVTeNbO$_x$ is, by far, the most selective catalyst towards acrylic acid and achieves a maximum of 71.3 % selectivity at 350 °C at propane conversion of 26.1%. MoVO$_x$, VPP and a-VPP reach significant, although lower, acrylic acid selectivities (7.2, 18.6, and 14.2 %, respectively). The maximum selectivity for MoVO$_x$ occurs at 225 °C. For VPP and a-VPP the optimal temperatures with respect to selectivity are 400 and 350 °C, respectively. This is in line with the fact that MoVO$_x$ is active at lower temperatures, while VPP and a-VPP require higher temperatures to convert propane (Fig. 2A). Finally, V$_2$O$_5$ achieves a much lower acrylic acid selectivity of 2.3%. The catalysts β-VOPO$_4$, α-VWOPO$_4$, α-VOPO$_4$ and VWPO$_x$ are unselective towards acrylic acid under the reaction conditions considered. The presence of such diverse scenarios in the dataset is crucial for the success of the AI approach.

**Table 1.** Catalyst properties and reaction parameters used as primary features for the SISSO analysis. The subscripts on the atomic compositions ($x$) and oxidation states ($\Omega$) indicate if the value concerns the bulk (b) or surface (s). The subscripts fr, and act indicate if the property concerns a fresh or activated sample, respectively. The subscript rxn,dry; rxn,wet; and rxn,C3 indicate properties measured with *in situ* NAP-XPS under dry, wet or $C_3$-rich gas-phase feeds, respectively.

| symbol | unit | description | technique |
|---|---|---|---|
| $T$ | °C | temperature (of reactivity measurement) | - |
| $V_{act}^{cell}$ | Å$^3$ | normalized unit cell volume | XRD (*ex situ*) |
| $s_{fr}, s_{act}$ | m$^2 \cdot$g$^{-1}$ | specific surface area | N$_2$ ads. (*ex situ*) |
| $V_{fr}^{pore}, V_{act}^{pore}$ | cm$^3 \cdot$g$^{-1}$ | pore volume | |
| $x_{b,fr}^{V}, x_{b,act}^{V}$ | % atom | bulk atomic content | XRF (*ex situ*) |
| $x_{s,fr}^{V}, x_{s,act}^{V}, x_{s,fr}^{O}, x_{s,act}^{O}, x_{s,fr}^{C}, x_{s,act}^{C}$ | % atom | surface atomic content | lab-XPS (*ex situ*) |
| $\Omega_{s,fr}^{V}, \Omega_{s,act}^{V}$ | e | oxidation state | |
| $\lambda^V, \lambda^O, \lambda^C$ | nm | inelastic mean free path | |
| $a_{fr}^{C-C}, a_{fr}^{C-O}, a_{fr}^{C=O}$ $a_{act}^{C-C}, a_{act}^{C-O}, a_{act}^{C=O}$ | % area | relative amount of carbon 1$s$ components | |
| $u_{m,fr}^{O_2}$ | μmol O$_2 \cdot$g$^{-1}$ | O$_2$ uptake per mass | TPRO (*ex situ*) |
| $u_{s,fr}^{O_2}$ | μmol O$_2 \cdot$m$^{-2}$ | O$_2$ uptake per surface area | |
| $x_{s,rxn,dry}^{V}, x_{s,rxn,dry}^{O}$ $x_{s,rxn,wet}^{V}, x_{s,rxn,wet}^{O}$ $x_{s,rxn,C3}^{V}, x_{s,rxn,C3}^{O}$ | % atom | surface composition | NAP-XPS (*in situ*) |
| $\Omega_{s,rxn,dry}^{V}, \Omega_{s,rxn,wet}^{V}, \Omega_{s,rxn,C3}^{V}$ | e | oxidation state | |
| $\lambda_{rxn}^V, \lambda_{rxn}^O$ | nm | inelastic mean free path | |
| $VB_{rxn,dry}, VB_{rxn,wet}, VB_{rxn,C3}$ | eV | valence band onset | |
| $W_{rxn,dry}, W_{rxn,wet}, W_{rxn,C3}$ | eV | work function | |
| $\sigma_{act}^{ref}$ | S·m$^{-1}$ | reference conductivity | MCPT (*in situ*) |
| $\Delta\sigma_{act}^{v}$ | S·m$^{-1}$ | conductivity stoichiometry-dependence | |
| $\Delta\sigma_{act}^{T}$ | S·ms$^{-1}$ | conductivity retention-time-dependence | |
| $\widetilde{\Delta\sigma}_{act}^{v}$ | % | $\Delta\sigma_{act}^{v}$ normalized by $\sigma_{act}^{ref}$ | |
| $\widetilde{\Delta\sigma}_{act}^{T}$ | %·s$^{-1}$ | $\Delta\sigma_{act}^{T}$ normalized by $\sigma_{act}^{ref}$ | |
| $E_{A,act}^{\sigma}$ | kJ·mol$^{-1}$ | activation energy of conductivity | |

The algorithm indeed needs to be informed about materials with different performance (in particular both desirable and undesirable types of behaviors) in order to identify the reactivity patterns we are searching for.

To gather information characterizing the catalysts and reflecting the potentially relevant processes governing selective oxidation within our AI approach, we measured a wide range of bulk and surface properties of both fresh and activated catalyst samples. We used the following common characterization techniques: x-ray diffraction (XRD), N$_2$ adsorption (ads.), x-ray fluorescence (XRF), laboratory x-ray photoelectron spectroscopy (lab-XPS), temperature-programmed reduction/oxidation (TPRO). Additionally, we measured properties of the activated catalyst samples under the reaction conditions (temperature and gas-phase feed) by the advanced techniques near-ambient-pressure XPS (NAP-XPS) and microwave cavity perturbation technique (MCPT). These advanced techniques, referred to as *in situ* (as opposed to the *ex situ* common techniques above), provide properties of the "working catalyst", which therefore take into account the catalyst dynamics. In particular, NAP-XPS, which provides surface properties, was carried out in three different feeds: dry, wet and $C_3$-rich. These conditions are used to probe the influence of surface composition and its electronic properties, which depends on the feed due to catalyst dynamics, on reactivity. Regarding MCPT, it is a technique for contactless determination of the conductivity, free of electrode effects.[25]

Finally, we note that microscopic as well as mesoscopic properties of the catalysts are included in our analysis, which can be related to phenomena at different length (and time) scales. For instance, the surface atomic composition (from lab-XPS and NAP-XPS) might characterize a molecular-level process whereas the pore volume (from N$_2$ ads.) may be associated to the diffusion of reactants and products on the catalyst pores, a transport phenomenon occurring at a larger length scale. The characterization of these catalysts, also performed following standardized protocols (described in ref. 12) represents an unprecedented effort to acquire a consistent and detailed set of more than fourty catalyst bulk and surface properties. An overview of the measured properties is shown in Table 1 (the full data set is available in ESI).

**AI approach**

The identification of correlations between materials and process properties on one side and the catalytic performance towards selective oxidation on the other was done by the SISSO approach.[13, 14] SISSO identifies *descriptors* in the form of typically complex, nonlinear analytical expressions depending on input parameters, called *primary features*. In the machine-learning nomenclature, these descriptors are representations. Thus, SISSO is also an efficient "representation-learning" algorithm.

The SISSO approach starts with the collection of its input parameters, called *primary features*. These include all possibly relevant physicochemical parameters that may relate to the processes governing the catalysis question of interest. Concerning these parameters, it is better to offer many possibilities, and it does not matter if some of these "primary features" are correlated with others. Our choice of primary features for this work is given in Table 1. They correspond to the measured materials properties as well as reaction parameters such as the temperature. Altogether, these are fifty. In the second step, we construct the descriptor candidates. For this, the primary features are systematically combined using mathematical operators such as addition, multiplication, difference, etc. (see details in ESI). Thus, we follow a symbolic regression approach.[26-28] This step results in the generation of *billions* of descriptor candidates. Each of them provides different numerical values for the different materials and/or processes. Thus, the big data challenge is related to the intricacy of the underlying physics and chemistry. From the large number of candidate descriptors, and using the provided values of the targets properties for the materials in the data set, SISSO selects very few, typically just $D$ = 1, or 2, or 3 best descriptor candidates, whose linear combination, with weighting coefficients, provides the best fit to the target property. $D$ is referred to as the descriptor dimension. The final selected descriptor is thus the vector containing the selected candidate descriptors as individual components. The selection of descriptors and the identification of the coefficients is done by compressed sensing.[29-31] The resulting models for the target property $P$ have the form

$$P^{(\text{SISSO})} = c_0 + \sum_{i=1}^{D} c_i d_i \text{ (eq. 4)},$$

where $d_i$ are the descriptor components selected from the many billions of candidates, and $c_i$ are the fitting coefficients. Importantly, only few primary features, out of the fifty offered ones, appear in the finally selected descriptor. The SISSO-derived descriptors are interpretable in the sense that one can identify the key primary features by simply inspecting the output expressions. These primary features are the relevant "catalyst genes".

Because the functional forms of the immensity of descriptor candidates offered to the SISSO analysis are very flexible, it is important to avoid *overfitting,* i.e., to avoid models that fit the provided data but are not generalizable. Two parameters control the model complexity: $D$ (see Eq. 4) and the number of times the mathematical operators are iteratively applied to the features (depth of symbolic-regression tree) in order to generate the descriptor candidates, called hereafter rung, and denoted by $q$. The model complexity is assessed using leave-one-material-out cross-validation (CV). This CV procedure consists of training models with a data set in which one of the catalysts is removed, and then using the so-obtained *ensemble* of best models to predict the property of the left-out material. This procedure is iterated until all the catalysts are left-out once. The root mean squared errors (RMSEs) averaged over all CV iterations (averaged CV-RMSEs) are used as our performance metric. The optimal complexity is considered the one with the lowest CV-RMSE. Further details on the CV procedure are provided in ESI.

In this work, we use the multi-task version of SISSO (MT-SISSO).[14]

In the context of SISSO, multi-task refers to a transfer-learning approach for the identification of a *single multi-dimensional descriptor* for a target property across different material classes or external conditions, each of them corresponding to *different fitting coefficients* ($c_i$ in Eq. 4). MT-SISSO thus provides a single descriptor for the property and different models for each class of materials or external condition. In the case of this work, such different external conditions correspond to the different reaction temperatures applied in the catalyst test. Predicting the target property at each of the measured temperatures is therefore a different *task*. We stress that, in addition to allowing for the simultaneous modelling of the catalytic performance at different temperatures, MT-SISSO also enables us to efficiently exploit the experimental data available, increasing the effective number of data points. This is because every material is measured in a large range of temperatures, providing multiple data points per material. The multi-task approach thus improves the reliability of the identified descriptors.

In spite of the application of MT-SISSO, the number of data points in our experimental data set is very small and by no means comparable to the amount of data needed for widely used machine-learning approaches as, e.g., kernel ridge regression or artificial neural networks. The latter typically requires >$10^3$ data points. However, we stress that for the SISSO approach a large amount of physicochemical information about the considered materials is provided by the immense number of descriptor candidates considered, with their quantitative values. The AI strategy of SISSO enables the identification of descriptors that capture the intricate underlying processes without the need for a large amount of experimentally characterized materials. The big-data aspect is thus in the intricacy of the materials functions as the signal to be reconstructed (using the language of compressed sensing). It does not focus on the number of materials (or observations). We nevertheless point out that the more experimental data are available, the more generalizable the SISSO-derived models will be. Obviously, SISSO can only capture processes that are governing the target properties in the employed experimental data.

**Identifying catalyst genes of the selective $C_3$-oxidation**

To identify descriptors indicating the catalyst performance in $C_3$-oxidation, we use the acrylic acid selectivity ($S_{\text{acrylic acid}}$) as target property. With the exception of MoVOx, which was only measured at 4 temperatures (*vide supra*), 9 different temperatures in the range 225-425 °C are considered per catalyst. Altogether, 76 data points are used. Even though the temperature is offered as a primary feature in our analysis and it can therefore be used to construct the descriptor expression, by using MT-SISSO we are also able to capture the effect of temperature via the coefficients used to fit the models ($c_i$ in Eq. 4). This is because such coefficients are functions of the task, in this case the different temperatures. Indeed, SISSO captures the temperature effect only by the fitted coefficients. The best descriptor expressions identified do not contain the temperature as a parameter (*vide infra*).

The errors obtained when $S_{\text{acrylic acid}}$ is estimated using the MT-SISSO model trained on the whole data set, i.e., the training errors, decrease as the rung $q$ and the dimension $D$ increase (dashed lines in Fig. 3). The training RMSEs are practically zero at $D = 3$ for the three considered $q$. This evidences the flexibility of expressions selected by SISSO to fit the input data. The average CV-RMSEs (solid lines in Fig. 3), however, do not decrease monotonically with rung and dimension. Instead, the average CV-RMSEs achieve a minimum value of 6.76% at $q = 3$, $D = 2$ with respect to an optimal predictability. This is therefore the identified appropriate complexity. The training error for such model is 1.46%. We note that a large fraction of the CV-RMSE is associated to the CV iteration in which the MoVTeNbO$_x$ catalyst is left-out. Since this catalyst achieves a much higher $S_\text{acrylic acid}$ compared to the remaining ones in the data set (Fig. 2B and Fig. 2C), it is probably dominated by a different process and therefore it is hard to correctly predict its performance based only on the remaining materials.

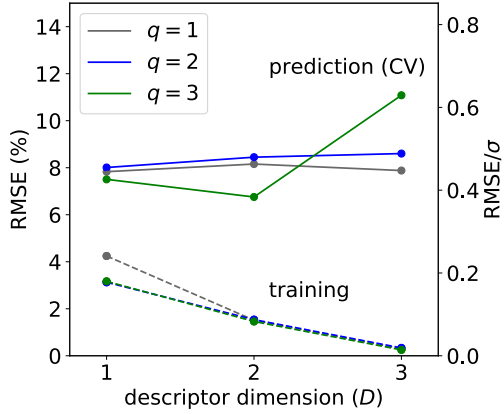

**Figure 3.** CV analysis of models derived by MT-SISSO for the acrylic acid selectivity. The CV errors shown correspond to the averaged RMSE across leave-one-material-out-CV iterations. The optimal complexity is $q = 3$, $D = 2$. The secondary axis (on the right) shows the CV-RMSE as a fraction of the standard deviation of $S_\text{acrylic acid}$ over the whole dataset. The CV-RMSE values shown here correspond to an ensemble size of 25 descriptors (see ESI for CV details).

The best descriptor identified by MT-SISSO, i.e., the descriptor identified using the whole data set at the optimal complexity provided the following model

$$S_\text{acrylic acid}^\text{(SISSO)}(T) = c_1^S(T)\left((V_\text{fr}^\text{pore})^2 \frac{W_\text{rxn,wet}}{E_\text{A,act}^\sigma} \frac{1}{(x_\text{s,rxn,C3}^\text{V} - x_\text{s,act}^\text{V})a_\text{act}^\text{C-O} x_\text{s,fr}^\text{C}}\right)$$

$$+ c_2^S(T)\left(V_\text{fr}^\text{pore} V_\text{act}^\text{pore} \frac{1}{E_\text{A,act}^\sigma} \frac{x_\text{s,rxn,wet}^\text{V}}{(x_\text{s,rxn,C3}^\text{V} - x_\text{s,act}^\text{V})(x_\text{s,rxn,dry}^\text{V} + x_\text{s,fr}^\text{C})}\right) \text{ (eq. 5),}$$

where the coefficients $c_1^S(T)$ and $c_2^S(T)$ depend on the measured temperature. In this expression, $V_\text{fr}^\text{pore}$ and $V_\text{act}^\text{pore}$ are the pore volumes of the fresh and activated catalysts, respectively, $E_\text{A,act}^\sigma$ is the activation energy of conductivity of the activated catalyst, $W_\text{rxn,wet}$ is the catalyst work function under reaction wet feed, $x_\text{s,act}^\text{V}$, $x_\text{s,rxn,dry}^\text{V}$, $x_\text{s,rxn,wet}^\text{V}$, and $x_\text{s,rxn,C3}^\text{V}$ are the V surface content of the activated catalyst and of the material under reaction dry, wet and C$_3$-rich feeds, respectively, $x_\text{s,fr}^\text{C}$ is the C surface content of the fresh catalyst, and $a_\text{act}^\text{C-O}$ is the fraction of surface carbon assigned to C-O in the activated catalyst. Figure 4A shows the model derived by MT-SISSO for $S_\text{acrylic acid}$ (eq. 5) evaluated for the materials and temperatures measured in the catalyst test (crosses) as well as the experimental measurements and indicates the good quality of the fit.

The SISSO-identified primary features are thus $V_\text{fr}^\text{pore}$, $V_\text{act}^\text{pore}$, $E_\text{A,act}^\sigma$, $W_\text{rxn,wet}$, $x_\text{s,act}^\text{V}$, $x_\text{s,rxn,dry}^\text{V}$, $x_\text{s,rxn,wet}^\text{V}$, $x_\text{s,rxn,C3}^\text{V}$, $x_\text{s,fr}^\text{C}$ and $a_\text{act}^\text{C-O}$. $V_\text{fr}^\text{pore}$ and $V_\text{act}^\text{pore}$ are associated to the porous structure of the catalyst and reflect processes related to the catalyst pores, for instance diffusion of reactants and/or products. $E_\text{A,act}^\sigma$ and $W_\text{rxn,wet}$ correspond to the activation energy of charge carrier transport and to the electronic surface potential at reaction conditions, respectively.

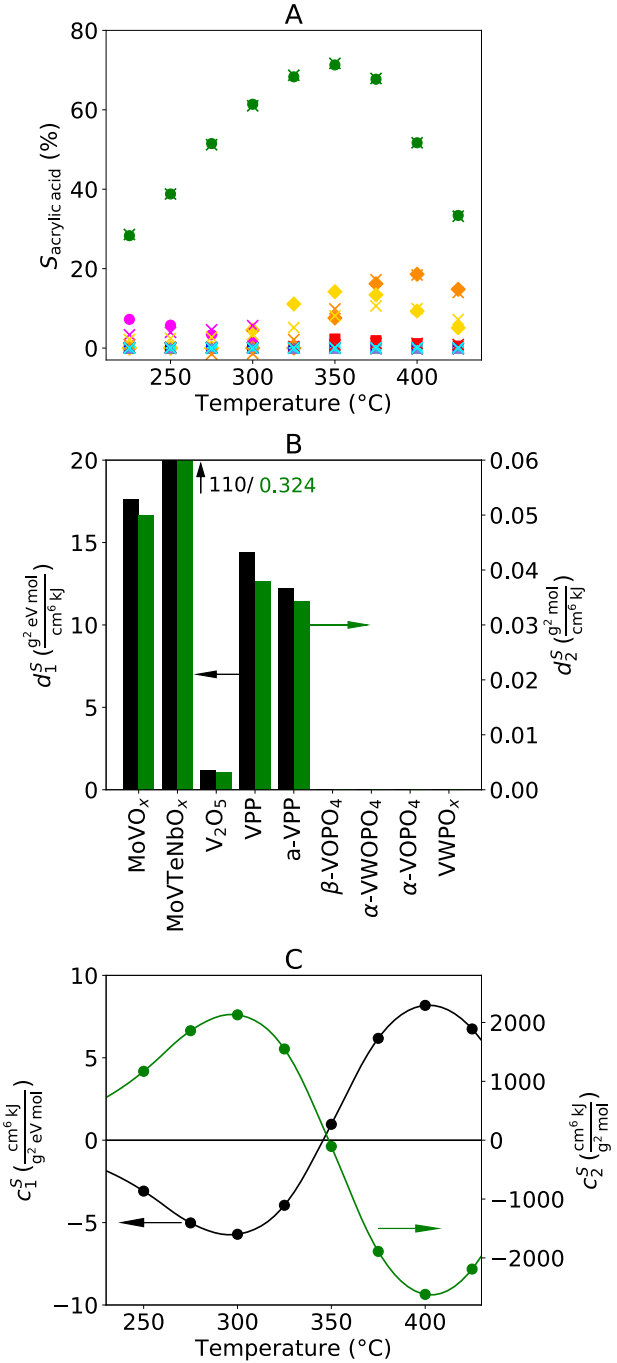

**Figure 4.** Descriptors identified by MT-SISSO for the acrylic acid selectivity ($S_\text{acrylic acid}(T)$). (A) Model expression evaluated on the nine vanadium-based catalysts of the data set at the measured temperatures (crosses), showing the quality of the fit with respect to experimental values (other markers). (B) Values of the best descriptor components for each catalyst. (C) Coefficients of the best model. The same markers and colors defining the materials in Fig. 2 are used in (A). The points in (C) are connected by splines (2$^\text{nd}$ order).

These primary features characterize electronic properties of the working catalysts, which can be related to the charge transfer from the catalyst to adsorbed reaction intermediates. $x_\text{s,act}^\text{V}$, $x_{s,\text{rxn,dry}}^V$, $x_{s,\text{rxn,wet}}^V$, and $x_{s,\text{rxn,C3}}^V$ indicate the relevance of the concentration of the redox-active element, vanadium, at the surface of the catalysts. Finally, $x_{s,\text{fr}}^C$ and $a_{\text{act}}^{C-O}$ are associated to the amount and specific types of surface carbon identified by XPS.

They reflect the strength of adsorption on specific catalyst surface sites and are thus related to surface site-specific processes. Altogether, the identified catalyst genes reflect a concerted and intricate interplay of catalyst bulk and surface processes that governs the selectivity towards acrylic acid. These include the catalyst dynamics, described by properties measured *in situ*, as well as transport phenomena at a higher length scale, encoded by the catalyst pore volume.

The key primary features (genes) identified in the SISSO analysis are consistent with previous investigations of selective oxidation catalysis in vanadium-based materials. In particular, in both descriptor components the difference $(x_{s,\text{rxn,C3}}^V - x_{s,\text{act}}^V)$ appears, which could be linked to the observed V surface enrichment occurring at reaction conditions in selective oxidation catalysts.[20-22] We note that the precise mathematical expression and the primary features individually contain less physical meaning than their collective interplay, since descriptors obtained with different mathematical operators or different primary features - for instance, correlated with those shown in Eq. 5 - can capture the same underlying processes.

The model identified by MT-SISSO (eq. 5) is based on a 2-dimensional (2-D) descriptor with components $d_1^S$ and $d_2^S$. These are different constant values for each material, and they are weighted by temperature-dependent coefficients. The descriptor components (Fig. 4B) assume non-negligible values for the catalysts that produce acrylic acid: MoVO$_x$, MoVTeNbO$_x$, V$_2$O$_5$, VPP, and a-VPP. Furthermore, their values are much higher for MoVTeNbO$_x$ compared to the other materials, in line with its much higher selectivity (Fig. 2B). For the catalysts β-VOPO$_4$, α-VWOPO$_4$, α-VOPO$_4$ and VWPO$_x$, both $d_1^S$ and $d_2^S$ are practically zero.

The coefficients $c_1^S(T)$ and $c_2^S(T)$ (in black and green, respectively, in Fig. 4C) of the acrylic acid selectivity take up positive and negative values depending on the temperature range. $c_1^S(T)$ and $c_2^S(T)$ are positive for high and low temperatures, respectively. The signs of the coefficients change at ca. 350 °C. Therefore, the selectivity is described by the model of eq. 5 as a sum of a positive and a negative term. This hints at different processes that facilitate and hinder the selectivity in a concerted and temperature-dependent manner. By fitting smooth functions to the coefficients, models for estimating acrylic acid selectivity across temperatures, including those not measured experimentally, can be obtained. Such models are useful for optimizing process conditions.

In addition to the selectivity, we also identified descriptors for the efficiency of propane oxidation, indicated by the propane conversion ($X_{\text{propane}}$). The SISSO-identified model, which corresponds to the optimal predictability of $q = 1$, $D = 2$ is

$$X_{\text{propane}}^{(\text{SISSO})}(T) = c_1^X(T)\left(\frac{u_{m,\text{fr}}^{O_2}}{W_{\text{rxn,wet}}}\right) + c_1^X(T)\left(u_{m,\text{fr}}^{O_2} x_{s,\text{rxn,C3}}^O\right) \text{ (eq. 6).}$$

In this expression, $u_{m,\text{fr}}^{O_2}$ is the reversible oxygen uptake of the fresh catalyst per mass, and $x_{s,\text{rxn,C3}}^O$ is the O surface content under reaction C$_3$-rich feed. $u_{s,\text{fr}}^{O_2}$ indicates the materials ability to reversibly incorporate oxygen on its bulk structure and is related, for instance, to the role of lattice oxygen. $W_{\text{rxn,wet}}$ and $x_{s,\text{rxn,C3}}^O$, in turn, are related to surface processes. While the $c_1^X(T)$ in the model of eq. 6 is positive for all considered temperatures, $c_2^X(T)$ is always negative (see Fig. S3C). This indicates that the processes captured by the first term in eq. 6 facilitate propane conversion whereas those associated to the second term hinder it.

Even though the 2-D descriptor in eq. 6 does reflect an interplay of processes governing activity, the descriptor complexity ($q = 1$) is lower compared to the case of acrylic acid selectivity (eq. 5, $q = 3$). Indeed, it is expected that the selectivity towards the oxygenate depends on a more intricate interplay of processes compared to the propane conversion to any product, i.e., including CO$_2$. We note that $W_{\text{rxn,wet}}$ is identified as a key parameter for both properties, consistent with the fact that the acrylic acid selectivity and the propane conversion are related (Fig. 2C) and might display some common governing process – and thus common materials genes.

By inspecting SISSO models predicting $S_{\text{acrylic acid}}$ and $X_{\text{propane}}$, we observe that the descriptor components $d_1$ and $d_2$ have large mutual linear correlation for the materials in the training dataset. This can be formalized by noticing that a linear model $d_2 = \alpha d_1 + \beta$ yields a good approximation of $d_2$ when $d_1$ is known. In other words, all the so- far known materials lie close to a straight line in the $(d_1, d_2)$ space. Since $d_1$ and $d_2$ depend on primary features that are measured in experiments for actual materials, it is unclear if materials that would land away from the $d_2 = \alpha d_1 + \beta$ line actually exist. However, both models are expected to become less reliable the further a new tested material lands from the $d_2 = \alpha d_1 + \beta$ line. Should this happen for a new tested material, the model would need re-training as a more complex model is likely needed. This can also be realized when noticing that the linear models $X_{\text{propane}}, S_{\text{acrylic acid}} = c_1 d_1 + c_2 d_2$ can predict values of $X_{\text{propane}}$ and $S_{\text{acrylic acid}}$ outside the physically meaningful interval 0-100%, *for arbitrary values of $d_1$ and $d_2$*, which are very different from the $(d_1, d_2)$ values that represent the materials in the dataset. We also notice that for $X_{\text{propane}}$, the model is particularly sensitive when $d_2$ departs from the $d_2^X = \alpha^X d_1^X + \beta^X$ line, which limits its applicability. A deeper analysis of this model will be published elsewhere.

**Maps of catalysts for guiding the design of new materials**

We used the relationship $d_2^S = \alpha^S d_1^S + \beta^S$ to obtain a "map of catalysts" (Fig. 5) showing $S_{\text{acrylic acid}}^{(\text{SISSO})}(T)$ as a function of two variables, the materials descriptor $d_1^S$ and the temperature. In the expression $d_2^S = \alpha^S d_1^S + \beta^S$, $\alpha^S$ and $\beta^S$ are fitted parameters, with values 2.948 10$^{-3}$ eV$^{-1}$ and -9.246 10$^{-4}$ g$^2$·mol·cm$^{-6}$·kJ$^{-1}$, respectively. The resulting map shows the selectivity, as a color scale from 0 to 100%, for the temperature range used in the experiment. Every material is represented, in this plot, by a horizontal line and the black lines indicate the materials in the experimental data set that produce acrylic acid and were used for training the model.

The map of Fig. 5 highlights the different types of behavior present in the data set used for the derivation of the descriptor. In particular, it shows that the temperature of maximum acrylic acid selectivity decreases as one moves from low to high $d_1^S$ values. The map also evidences the unique and higher performance of MoVTeNbO$_x$ compared to the other materials. Additionally, this materials chart can accelerate the design of new catalysts, since it indicates the regions of the materials space where high-performant materials are found. In particular, catalysts with high $d_1^S$ values (regions shown in blue in Fig. 5) are associated to high selectivity towards the formation of acrylic acid. From eq. 5, the pore volumes and the activation energy of conductivity are the key properties that impact the value of $d_1^S$. By increasing pore volumes and decreasing activation energy of conductivity, for instance, $d_1^S$ is increased. To illustrate how materials with "better" parameters will exhibit a better selectivity, we imagine a hypothetical catalyst, which would be obtained from VPP, increasing its pore volumes by 50 % (in both fresh and activated materials) and decreasing its activation energy of conductivity by 50 %. The VPP$_{modified}$ material, shown as the green dashed line in Fig. 5, would provide an acrylic acid selectivity ca. 3 times higher than that for VPP. For this gedankenexperiment, we assume that all other primary features of the catalyst remain unchanged.

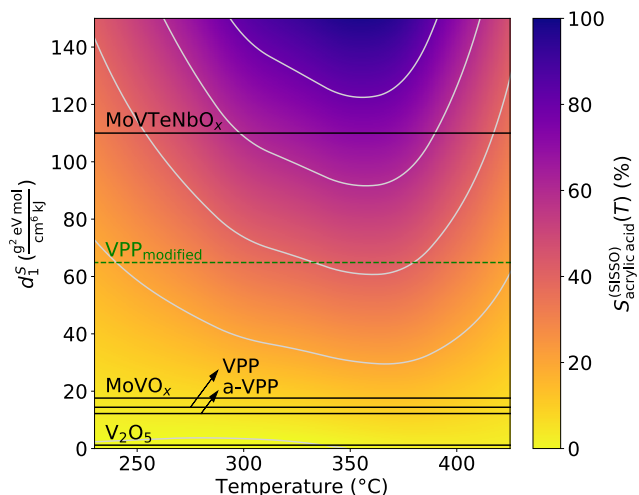

**Figure 5.** Map of catalysts given by the MT-SISSO model for the acrylic acid selectivity $S_{\text{acrylic acid}}^{(\text{SISSO})}(T)$, indicating the regions of the materials space corresponding to high selectivity (in blue). The materials used for deriving the descriptors that produce acrylic acid are indicated by the black lines. The green dashed line indicates a hypothetical VPP$_{modified}$ catalyst. VPP$_{modified}$ would be obtained by increasing 50% VPP's pore volumes (in both fresh and activated materials) and by decreasing 50% its activation energy of conductivity. The splines shown in Fig. 4C are used to interpolate $c_1^S(T)$ and $c_2^S(T)$ across temperatures.

Finally, we note that the models derived by MT-SISSO are expected to hold for catalysts obtained as modification of the vanadium-based catalysts as well for other (new) materials as long as the reactivity is governed by the same processes governing the catalytic performance in the nine materials of our data set. However, for regions of the materials space containing catalysts very different from those in the data set, the MT-SISSO model might need to be re-trained with more data. This calls for a systematic experimental exploration of the materials space, i.e., addition information at $(d_1, d_2)$ regions whose experimental data are missing, at present.

## Conclusions

Our study shows how consistent data in heterogeneous-catalysis research, generated according to standardized protocols for performing and annotating experiments,[12] enable the identification of the key descriptive parameters related to catalyst performance, the "materials genes of catalysis", by AI. Nine vanadium-based alkane oxidation catalysts presenting diverse reactivity towards C$_3$-oxidation were synthesized, characterized, and tested according to such procedures. In particular, their detailed characterization resulted in more than forty measured properties per material. To such data set, presenting a small number of materials but a very large amount of information for each catalyst, provided in terms of input features, we applied the compressed-sensing symbolic-regression SISSO approach. Out of billions of descriptor candidates, we found nontrivial interpretable expressions reflecting the concerted interplay of processes that govern catalysis, including the crucial catalyst dynamics. The AI-identified descriptors enable us to generate maps of catalysts for guiding the search of novel materials and rationalizing the reactivity trends. In particular, the key catalysts properties related to acrylic acid selectivity include the pore volume, the activation energy of conductivity, the work function, the fraction of surface carbon species assigned to carbon-oxygen as well as the V and C surface contents. These properties, measured by N$_2$ adsorption, *in situ* MCPT and XPS (including *in situ* NAP-XPS), are thus the key ones to be measured and used for the design of selective materials.

The combination of systematic experiments and AI proposed here is suitable for improved materials discovery and the modelling of complex materials properties and functions whose underlying governing processes are intricate and hard to model explicitly by atomistic simulations.

## Supplementary material

SISSO and CV details as well as additional results for the descriptor $X_{\text{propane}}^{(\text{SISSO})}(T)$ are available in ESI.

The SISSO analysis described in this publication can be found in a Jupyter notebook at the *NOMAD Artificial-Intelligence Toolkit* (https://nomad-lab.eu/AIToolkit/), where it can be repeated and modified directly in a web browser.

## Author information

### Corresponding author

*foppa@fhi-berlin.mpg.de

### ORCID IDs

Lucas Foppa: https://orcid.org/0000-0003-3002-062X
Luca M. Ghiringhelli: https://orcid.org/0000-0001-5099-3029
Annette Trunschke: https://orcid.org/0000-0003-2869-0181

### Current addresses

†Peter Kraus: School of Molecular and Life Sciences, Curtin University, GPO Box U1987, Perth 6845, WA, Australia.

## Acknowledgments

L. F. acknowledges the funding from the Swiss National Science Foundation, postdoc mobility grant P2EZP2_181617 and the NOMAD CoE (European Union's Horizon 2020 research and innovation program under the grant agreement N° 951786). Dr. Thomas A. R. Purcell is acknowledged for providing the SISSO++ code. Stephen Lohr and Sven Richter are acknowledged for the synthesis of the materials, Ezgi Erden for assisting the catalyst test, Dr. Olaf Timpe for chemical analysis, Dr. Gregory Huff, Dr. Toyin Omojola, Dr. Yuanqing Wang, Dr. Jinhu Dong, Dr. Gregor Koch, and Dr. Detre Teschner for their help in performing the NAP-XPS experiments. The Helmholtz-Zentrum Berlin is acknowledged for providing beamtime and for the continuous support of the ambient pressure XPS activities of the MPG at BESSY II. This work was conducted in the framework of the BasCat collaboration between BASF SE, Technical University Berlin, Fritz-Haber-Institut der Max-Planck-Gesellschaft, and the clusters of excellence "Unified Concepts in Catalysis"/"Unifying

Systems in Catalysis" (UniCat https://www.unicat.tu-berlin.de/UniSysCat https://www.unisyscat.de).